\documentclass[11pt]{article}
\input{psfig.sty}
\begin{document}

\title{
\textbf{Surface tension and interfacial fluctuations\\ 
in $d$--dimensional Ising model}
}

\author{J. Kaupu\v{z}s
\thanks{E--mail: \texttt{kaupuzs@latnet.lv}} \\
Institute of Mathematics and Computer Science, University of Latvia\\
29 Rainja Boulevard, LV--1459 Riga, Latvia}

\date{\today}

\maketitle

\begin{abstract}
The surface tension of rough interfaces between coexisting phases
in 2D and 3D Ising models are discussed in view of the known
results and some original calculations presented in this paper.
The results are summarised in a formula, which allows to interpolate the
corrections to finite--size scaling between two and three dimensions.
The physical meaning of an analytic continuation to noninteger values
of the spatial dimensionality $d$ is discussed.
Lattices and interfaces with properly defined fractal
dimensions should fulfil certain requirements to possibly
have properties of an analytic continuation from
$d$--dimensional hypercubes. Here $2$
appears as the marginal value of $d$ below which
the $(d-1)$--dimensional interface splits in disconnected
pieces. Some phenomenological arguments are proposed
to describe such interfaces. They show that the character
of the interfacial fluctuations at $d<2$ is not the same
as provided by a formal analytic continuation from
$d$--dimensional hypercubes with $d \ge 2$. 
It, probably, is true also for the related critical exponents.
\end{abstract}


\vspace*{1ex}


\section{Introduction}

The phase coexistence and surface tension of fluctuating interfaces
is an object of extensive theoretical studies. It covers exact results
for two--dimensional Ising and solid--on--solid (SOS) models~\cite{SPA},
low--temperature series analysis of 3D Ising model~\cite{SF},
studies within the cappilary wave approximation at arbitrary spatial
dimension $d$~\cite{GF}, a general phenomenological
description~\cite{Privman}, as well as Monte Carlo studies
of surface tension in 3D Ising model~\cite{MWLB,HP}.
We recommend the review papers~\cite{PrivRev,HasRev}
for further references.

 In this paper first we will briefly discuss the existing results
for rough interfaces in 2D and 3D Ising models, completing them by
some original calculations.
Then we will discuss the spatial dimensionality $d$ as a continuous
parameter from a purely formal point of view, as well as 
linking noninteger $d$ values to lattices
with certain fractal dimension. The latter consideration
suggests that 2 is a special marginal value of $d$.

\section{Surface tension of 2D Ising model}

Consider the Ising model with the Hamiltonian $H$,
\begin{equation}
\frac{H}{k_B T} = - \beta \sum\limits_{\langle ij \rangle} s_i s_j \;,
\end{equation}
where $k_B$ is the Boltzmann constant, $T$ is the temperature,
$s_i = \pm 1$ are the spin variables, and
$\beta$ is the coupling constant describing the ferromagnetic
interaction between all pairs $\langle ij \rangle$ of the
neighbouring spins. In the low--temperature phase
at $\beta>\beta_c$, where
$\beta_c= \frac{1}{2} \ln \left( 1+ \sqrt{2} \right)$ is the critical
coupling, certain interfacial structure can be imposed by appropriate
boundary conditions.

The surface tension of inclined interfaces in 2D Ising model
has been considered in~\cite{SPA,Privman}.
As defined in~\cite{SPA}, the spins are located at lattice points
$x=0,1,2,\ldots,L$ and
$y= \pm \frac{1}{2}, \pm \frac{3}{2}, \ldots,
\pm \left( M- \frac{1}{2} \right)$.
The interface which makes a mean angle $\theta$ with $x$ axis is forced
by the boundary conditions (see Fig.~1 in~\cite{SPA,Privman})
$s(x,\pm (M-1/2)) = \pm 1$, $s(0,y>0)=1$, $s(0,y<0)=-1$,
$s(L,y>m)=1$, $s(L,y<m)=-1$.
Hence, the endpoints of the interface
are pinned at $x=0; y=0$ and $x=L; y=m$, and $\tan \theta =m/L$.
The quantity of interest is the partition function $Z(m,L;M)$ of the
lattice with such an interface, normalized to the partition function
of the lattice without the interface, for which all the boundary spins
are fixed positive. 

The surface tension $\sigma(\theta,L;M)$ is defined as~\cite{SPA}
\begin{equation}
\sigma(\theta,L;M) = - \frac{\cos \theta}{L} \ln Z(L \tan \theta,L;M) \;.
\end{equation}
The bulk surface tension at $\theta=0$, i.~e.,
$\tau \equiv \sigma(0,\infty,\infty)>0$ and the surface stiffness
$\kappa>0$ are defined according to
\begin{equation}
\frac{\sigma(\theta,\infty,\infty)}{\cos \theta}
= \tau + \frac{1}{2} \kappa \theta^2 + O \left(\theta^4 \right) \;.
\end{equation}

The influence of the size $M$ decreases exponentially at $L \to \infty$
when $M=O(L)$~\cite{SPA}. Based on exact formula for
$Z(m,L) \equiv Z(m,L;\infty)$, it has been found in~\cite{SPA} that
\begin{equation}
Z(m,L) \simeq \exp \left( - \tau_L L \right)
\left( \frac{ \kappa_L}{2 \pi L} \right)^{1/2}
\exp \left( \frac{- \kappa_L m^2}{2L} \right)
\label{eq:ZmL}
\end{equation}
holds for large $L$ with
\begin{equation}
\tau_L = \tau + \frac{a}{L} + o \left( \frac{1}{L} \right),
\hspace{3ex}
\kappa_L = \kappa + \frac{b}{L} + o \left( \frac{1}{L} \right) \;.
\end{equation}
Here $\tau= 2 \, ( \beta - \beta^*)$,
where $\beta^*$ is the dual coupling
defined by $\exp(-2 \beta) = \tanh \beta^*$, 
$\kappa= \sinh \tau$, $a=2 \beta$,
and $b= \frac{1}{2} \left[ \sinh^2 2\beta^* + \sinh^2 \tau
+ 3 \cosh \tau \right]$.
The square root term in~(\ref{eq:ZmL}) is the normalization
factor of the Gaussian distribution.

Hence the surface tension $\tilde \sigma(0,L) \equiv \sigma(0,L;\infty)$
is~\cite{SPA}
\begin{equation}
\tilde \sigma(0,L) = \tilde \sigma(0,\infty) + \frac{\ln L}{2L}
+ \frac{a - \ln \left[(\kappa/2 \pi)^{1/2} \right]}{L} +
o \left( \frac{1}{L} \right) \;.
\label{eq:sigma}
\end{equation}
Note that the universal logarithmic correction term $\frac{1}{2}\ln L/L$
comes from the normalization factor in~(\ref{eq:ZmL}).
Similar corrections to scaling appear also in the case, discussed below,
where the interface is forced by anti--periodic boundary
conditions along one of the axes, the boundary conditions
being periodic along the other axis.

An exact expression for the partition function of a finite--size
2D lattice on a torus with arbitrary coupling constants between 
each pair of neighbouring spins has been reported in~\cite{Bednorz} 
obtained by the loop counting method and represented
by determinants of certain transfer matrices. In the standard 2D 
Ising model with only one common coupling constant $\beta$ these 
matrices can be diagonalized easily, using the standard 
techniques~\cite{Landau}. Besides, the loop counting method
can be trivially extended to the cases with antiperiodic
or mixed boundary conditions. It is necessary only to mention that
each loop gets an additional factor $-1$ when it winds round
the torus with antiperiodic boundary conditions. 
We consider the partition functions
$Z_{pp}$, $Z_{aa}$, $Z_{ap}$, $Z_{pa}$.
In this notation the first index refers to 
$x$ axis, and the second one -- to $y$ axis;
$p$ means periodic and $a$ -- antiperiodic boundary conditions.
Thus, for the lattice with $x=1,2,\ldots,N$ and $y=1,2,\ldots,L$,
we obtain the following exact expressions:
\begin{eqnarray}
Z_{pp} &=& \left( Q_1+Q_2+Q_3-Q_0 \right)/ \, 2 \nonumber \\
Z_{ap} &=& \left( Q_0+Q_1+Q_3-Q_2 \right)/ \, 2 \nonumber \\
Z_{pa} &=& \left( Q_0+Q_1+Q_2-Q_3 \right)/ \, 2 \label{eq:zz} \\
Z_{aa} &=& \left( Q_0+Q_2+Q_3-Q_1 \right)/ \, 2 \nonumber 
\end{eqnarray}
where $Q_0$ is the partition function represented by the sum of 
the closed loops on the lattice, 
as consistent with the loop counting method in~\cite{Landau}, 
whereas  $Q_1$, $Q_2$, and $Q_3$
are modified sums with additional factors
$\exp(\Delta x \cdot i \pi/N + \Delta y \cdot i \pi/L)$, 
$\exp(\Delta x \cdot i \pi/N)$, and 
$\exp(\Delta y \cdot i \pi/L)$, respectively, 
related to each change of coordinate $x$ by $\Delta x = \pm 1$,
or coordinate $y$ by $\Delta y = \pm 1$ when making a loop. 
The standard manipulations~\cite{Landau} yield
\begin{eqnarray} \label{eq:qi}
&&Q_i = 2^{NL} \prod\limits_{q_x, \, q_y} \left[ \cosh^2 (2 \beta)
-\sinh(2 \beta)  \right. \\
&&\left. \times \left( \cos \left[ q_x+ 
\left( \delta_{i,1}+\delta_{i,2} \right) \frac{\pi}{N} \right]
+ \cos \left[ q_y+ \left( \delta_{i,1}+\delta_{i,3} \right) 
\frac{\pi}{L} \right] \right) \right]^{1/2} \nonumber \;,
\end{eqnarray}
where the wave vectors $q_x=(2 \pi/N) \cdot n$ and 
$q_y=(2 \pi/L) \cdot \ell $ run over all the values
corresponding to $n=0, 1, 2, \ldots , N-1$
and  $\ell=0, 1, 2, \ldots , L-1$.
Eq.~(\ref{eq:qi}) represents an analytic extension
from small $\beta$ region~\cite{Bednorz}. The correct sign
of square roots is defined by this condition, and 
all $Q_i$ are positive except for $Q_0$, which
vanishes at $\beta=\beta_c$ and becomes negative at $\beta>\beta_c$.
The sign--alternating factor 
with $q_x=q_y=0$ can be written as $1- \sinh(2 \beta)$.
In the case of the periodic boundary conditions,
each loop of $Q_0$ has the sign $(-1)^{m+ab+a+b}$~\cite{Bednorz},
where $m$ is the number of intersections, $a$ is
the number of windings around the torus in $x$ direction, and 
$b$ -- in $y$ direction. The correct result for $Z_{pp}$ 
is obtained if each of the loops has the sign $(-1)^m$.
Eq.~(\ref{eq:zz}) for $Z_{pp}$ is then obtained
by finding such a linear combination of quantities $Q_i$ which ensures
the correct weight for each kind of loops.
Eqs.~(\ref{eq:zz}) for $Z_{aa}$, $Z_{ap}$, and $Z_{pa}$
are obtained in an analogous way.

The surface tension is given by the Onsager's ans\"atz 
\begin{equation}
\sigma(L,N) = L^{-1} \ln \left( N Z_{pp}/Z_{ap} \right) \;,
\end{equation}
where the size $N$ in $x$ direction is included since,
due to the
translation symmetry, each interface configuration has $N$ equivalent
copies obtained by shifting along the $x$ axis. It means that we take
only one of the $N$ equivalent copies for the interfacial partition
function $N^{-1} Z_{ap}/Z_{pp}$.

We have analysed the corrections to scaling for $\sigma(L,N)$
numerically.
Considering a trial function of the form
\begin{equation}
\sigma(L,L) \simeq \tau + A \ln L/L + B/L
\label{eq:sigap}
\end{equation}
with $\tau = 2 \, (\beta-\beta^*)$,
the coefficients $A$ and $B$ have been evaluated by fitting 
the calculated values of $\sigma(L,L)$ and $\sigma(2L,2L)$.
We have observed that the obtained effective
coefficient $A(L)$ converges almost
linearly in $1/L$ to certain asymptotic value,
whereas $B(L)$ plot looks more linear in the scale of $\ln L/L$.
It means that
\begin{eqnarray}
\label{eq:AB}
A(L) &\simeq& A(\infty) + C_A/L \;, \\
B(L) &\simeq& B(\infty) + C_B \ln L/L \nonumber
\end{eqnarray}
hold with some constants $C_A$ and $C_B$.
Our values of $A(L)$, computed at $\beta=0.5$, are $A(8) \simeq 0.62206$,
$A(16) \simeq 0.56727$, $A(24) \simeq 0.54638$, $A(32) \simeq 0.53546$,
and $A(40) \simeq 0.52876$. The corresponding values of $B(L)$ are
$1.21819$, $1.37012$, $1.43998$, $1.48036$, and $1.50687$.
We have extracted from these numbers the following estimates:
$A=A(\infty)=0.502 \pm 0.011$, $C_A=1.07 \pm 0.20$,
$B=B(\infty)=1.659 \pm 0.015$, and $C_B=-1.65 \pm 0.11$.
These values have been obtained by fitting $A(L)$ and $B(L)$
to~(\ref{eq:AB}) at $L=32, 40$, and the discrepancies between the estimates
at $L=32, 40$ and $L=8, 16$ have been assumed as the error bars,
indicating the range of possible deviations from
the true asymptotic values. According to the observed monotonous
behaviour of the coefficients, estimated from~(\ref{eq:AB}) for each
pair of sizes $L$ and $L-8$, these deviations, most probably, are
positive for $A$ and $B$ and negative for $C_A$ and $C_B$.
Our calculations show that $A$, likely, has the same universal value
$1/2$ as the coefficient at $\ln L/L$ in~(\ref{eq:sigma}), whereas
$B$ differs from the corresponding coefficient in~(\ref{eq:sigma}),
the latter being $B' \simeq 2.653678$ at $\beta=0.5$.
Besides, the observed deviations from the asymptotic law~(\ref{eq:sigap})
are characterised by a remainder term $\sim \ln L/L^2$,
which is compensated by $A \to A(L)$ and $B \to B(L)$.
Assuming that $A(\infty)=1/2$, the estimation of coefficient $C_A$
can be improved. From the first equation of~(\ref{eq:AB}), then we
obtain $C_A = 1.15 \pm 0.08$ at $L=40$. Here the discrepancy
with our previous value has been put for the error bars.

These calculations for system sizes up to $2L=80$ have been performed
by double--precision FORTRAN codes. In this case computations at
larger system sizes become problematic due to the rounding errors:
it is necessary for calculation of $Z_{ap}$ to
extract from the linear combination of $Q_i$ a quantity, which is
exponentially small relative to $\mid Q_i \mid$. Therefore, a more
precise estimation of the asymptotic values requires a computation with
substantially larger number of digits.

\section{Surface tension of 3D Ising model}
\label{sec:3d}

According to the phenomenological description provided in~\cite{Privman},
a relation similar to~(\ref{eq:ZmL}) holds for rough interfaces
(above the roughening transition temperature $T_R$
and below the bulk critical temperature $T_c$) also in three dimensions.
Note that the interface is always rough, i.~e., not pinned by the
underlaying lattice structure, in two dimensions
at nonzero temperature~\cite{Privman,PrivRev}.

Consider a $d$-dimensional ($d=2,3$)
$N \times L^{d-1}$ lattice, where $N$ is the linear size in the
direction perpendicular to the interface (when $\theta=0$).
In analogy to~\cite{Privman}, 
\begin{eqnarray}
Z(m,L) &\simeq& \exp \left( - \tau_L L^{d-1} \right)
R(L,d)
\exp \left( - L^{d-1} \kappa_L \theta^2 \right) \nonumber \\
&\simeq& \exp \left( - \tau_L L^{d-1} \right)
R(L,d)
\exp \left( - L^{d-3} \kappa_L m^2 \right)
\label{eq:Zd}
\end{eqnarray}
is expected for the partition function $Z(m,L)$ of an inclined interface
with small tilt angle $\theta \simeq m/L$, large $L$, and $N=O(L)$,
where $\tau_L$ and $\kappa_L$ are the finite--size
observables of the bulk surface tension $\tau$ and the stiffness
coefficient $\kappa$. According to the arguments provided
in~\cite{Privman}, $R(L,d)$ should behave like
the normalization factor of the Gaussian distribution, i.~e.,
\begin{equation}
R(L,d) \sim \left( L^{d-3} \kappa_L /2 \pi \right)^{1/2}
\label{eq:R}
\end{equation}
should hold for small values of $L^{d-3} \kappa_L /2 \pi$.
Eqs.~(\ref{eq:Zd}) and~(\ref{eq:R}) coincide with~(\ref{eq:ZmL})
at $d=2$ and with Monte Carlo (MC) simulation results for
tilted interfaces in three dimensions~\cite{MWLB}.

If the interface is forced by antiperiodic boundary
conditions along one of the axes (where the size is $N$) and periodic
boundary conditions along the other axes, then its mean slope
is zero, therefore the surface tension
\begin{equation}
\sigma(L,N) = L^{1-d} \ln \left( N Z_{pp}/Z_{ap} \right) \;,
\label{eq:sigd}
\end{equation}
should be more or less consistent with $\sigma = -L^{1-d} \ln Z(0,L)$
calculated from Eqs.~(\ref{eq:Zd}) and~(\ref{eq:R})
at $\theta=0$. The partition functions $Z_{pp}$ and $Z_{ap}$
in~(\ref{eq:sigd}) have the same meaning as before, only the second
index now refers to all axes aligned parallel to the interface.

The surface tension~(\ref{eq:sigd}) in 3D case has been properly
studied by Monte Carlo simulations in~\cite{HP} by means of the
thermodynamical integration of the interfacial energy.
It has been found that the surface free energy $F_s=L^2 \sigma(L,N)$
is well described by the expression of the Gaussian capillary
wave theory~\cite{GF}
\begin{equation}
F_s \simeq C_s + \sigma L^2 \;,
\label{eq:cap}
\end{equation}
where $C_s = G - \frac{1}{2} \ln \sigma$ with $G \approx 0.29$
holds near the critical point. Eq.~(\ref{eq:cap}) is consistent with
(\ref{eq:Zd}) and~(\ref{eq:R}), where the $\frac{1}{2} \ln \sigma$
term comes from~(\ref{eq:R}), taking into account that
$\kappa \propto \sigma$ holds at $\beta \to \beta_c$.
Contrary to the 2D case, now the leading correction
to scaling for $\sigma$ is $\sim 1/L^2$, as consistent with
$\tau_L = \tau + O \left( 1/L^2 \right)$, and the logarithmic
correction is absent since $d-3=0$ vanishes in~(\ref{eq:R}).

The ``endpoint'' correction of order $O(1/L)$ is expected in the case of
inclined interfaces (in 3D lattice) considered in~\cite{Privman}
due to the direct influence of the fixed boundary spins.
In the case of~(\ref{eq:sigd}) such a correction apparently
is absent according to~\cite{HP}. We have verified also
via direct simulation of the partition
functions $Z_{pp}$ and $Z_{ap}$ by the multicanonical Monte Carlo 
sampling method~\cite{Berg}
that $\sigma(L,L)$ at $\beta=0.3$ well coincides with
$\sigma(L,L) = \sigma(\infty,\infty) + C_s/L^2$ law within
$L \in [6;16]$.

\section{A formal generalisation to continuous dimension $d$}
\label{sec:fcon}

The relation~(\ref{eq:Zd}) is quite general and has to be
true for any natural $d \ge 2$ to provide finite values of
bulk surface tension and stiffness,
the only question is about the specific form of prefactor
$R(L,d)$~\cite{Privman}. Besides, it is possible to consider
the spatial dimension $d$ in~(\ref{eq:Zd}) and~(\ref{eq:R})
as a continuous parameter within $1<d \le 3$. The $d=3$ case
is marginal for the normalization factor~(\ref{eq:R}),
since the width of the distribution over $m$ in~(\ref{eq:Zd})
is diverging in the thermodynamic limit at $d<3$ and
becomes finite at $d=3$.
Due to the latter fact, Eq.~(\ref{eq:R}) at $d=3$, likely,
is valid only in vicinity of the bulk critical point,
where the distribution width is large.

An approximation for the case where the interface is forced by the mixed
boundary conditions (antiperiodic in one direction, periodic -
in other directions) is obtained by setting $\theta = 0$.
The MC results discussed in Sec.~\ref{sec:3d} suggest that
for this kind of boundary conditions the
corrections to scaling of the kind $1/L$, which appear in two dimensions,
have to be dechipered in general as $1/S$ corrections,
where $S = L^{d-1}$ is the interface area.
Thus, the surface tension $\sigma(L,N)$ of
the $N \times L^{d-1}$ lattice with $N=O(L)$ is expected to be
\begin{equation}
\sigma(L,N) = \sigma(\infty,\infty) + \frac{3-d}{2} \, \frac{\ln L}{L^{d-1}}
+ O \left( L^{1-d} \right) 
\label{eq:cont}
\end{equation}
for $d \le 3$. It allows to interpolate between two and three
dimensions. Our further consideration
shows that the continuation below $d=2$ is problematic,
if one tries to relate it to real physical systems.

\section{Physical interpretation of continuous dimension $d$}
\label{sec:phin}

To give some physical meaning to~(\ref{eq:Zd}), (\ref{eq:R}),
and~(\ref{eq:cont}) at a noninteger
$d$, one has to relate these formulae to some really existing lattices.
We will consider lattices with suitably defined
fractal dimension like in~\cite{TKHT}.
Such lattices with interfaces between the coexisting phases, probably,
should meet a lot of requirements to be considered in some sense
as analytic continuations from natural $d$.

In our further consideration it is suitable
to define the interface of a given spin configuration as a set of
interfacial spins located near the phase--separation border.
We denote by $\Lambda_S$ the subset of lattice sites where these spins
are located. For an arbitrary lattice, we consider the graph--theoretic
distance $\mbox{dist}(x,y)$ between the sites $x$ and $y$, which
is defined as the minimum number of bonds in $\Lambda_b$ that one
needs to connect $x$ and $y$. Here $\Lambda_b$ is the set of
bonds between the directly interacting neighbouring spins.
We denote by $N_R(x)$ the number of lattice sites
inside a sphere of radius $R$ centered at $x$, i.~e., the number
of those sites $y$ for which $\mbox{dist}(x,y) < R$ holds.
In analogy, $n_R(x)$ is defined as the number of interfacial sites
$y \in \Lambda_S$ inside the sphere of radius $R$ centered at
$x \in \Lambda_S$.

In the thermodynamic limit the $d$--dimensional hypercubes
($d=2, 3, 4, \ldots$) with $(d-1)$--dimensional interfaces have 
certain essential properties, listed below.
\begin{itemize}
\item[(i)]
The number of bonds in $\Lambda_b$ connected to one
lattice site is bounded uniformly (for all lattice sites)
from above by some positive constant.
\item[(ii)]
The lattice has certain dimension $d$, defined as
\begin{equation}
d = \lim\limits_{R \to \infty} \frac{\ln N_R(x)}{\ln R} \;,
\end{equation}
which holds for all $x$.
\item[(iii)]
The interface has certain dimension $d_s$, defined as
\begin{equation}
d_s = \lim\limits_{R \to \infty} \frac{\ln n_R(x)}{\ln R} \;,
\label{eq:ds}
\end{equation}
which holds for all $x \in \Lambda_S$, and this dimension
is equal to $d-1$.
\item[(iv)]
By any physically senseful definition of the set of interfacial
sites $\Lambda_S$, there exists a subset $\Lambda_s \subseteq \Lambda_S$
of these sites, which forms an infinitely large connected cluster.
In other words, the main body of the interface is connected.
\end{itemize}
There is no reason to expect that lattices and interfaces
with noninteger fractal dimension have properties of an analytic
continuation from the hypercubes with integer $d$ if any of these
requirements is violated.

The following lemma is relevant for our further considerations.
\vspace{2ex}

\textit{Lemma} -- If the interface has certain dimension $d_s$ such
that~(\ref{eq:ds}) holds for all $x \in \Lambda_S$, and
there exists a subset $\Lambda_s \subseteq \Lambda_S$ of the interfacial
sites which form an infinitely large connected cluster, then $d_s \ge 1$.
\vspace{2ex}

\textit{Proof}. \hspace{1ex} Choose $x, y \in \Lambda_s$ at a
distance $\mbox{dist}(x,y)=R$. By definition
of connected cluster, there exists a path connecting $x$ and $y$
by bonds of $\Lambda_b$ such that all sites of this path $y_i$
with $i=0,1,2, \ldots$, where $y_0 \equiv x$, belong to $\Lambda_s$.
Obviously, $\mbox{dist}(x,y_i)$ reaches $R$ for the first time at
some $i=i_0$.
By definition of the distance, $\mbox{dist}(x,y_i) \le i$ holds,
so that $i_0 \ge R$. Thus, there exists at least
$R$ sites $y_i \in \Lambda_s$ with $i=0,1,2, \ldots, i_0-1$
such that $\mbox{dist}(x,y_i) < R$, i.~e., $n_R(x) \ge R$ holds.
Since the cluster is infinitely large, we can choose unlimitedly
large $R$. Hence
\[
d_s = \lim\limits_{R \to \infty} \frac{\ln n_R(x)}{\ln R}
\ge \lim\limits_{R \to \infty} \frac{\ln R}{\ln R} = 1 \;,
\]
which proves the lemma.

According to this lemma, $d=2$ is the lower marginal value of
dimension $d$ at which properties (ii) to (iv) still
can be satisfied simultaneously. Hence, at $d<2$ the $(d-1)$--dimensional
interface cannot contain infinitely large connected clusters,
i.~e., it splits in disconnected finite--size pieces.
Thus, if we would choose $x \in \Lambda_S$ and look for the interfacial
structure within a sphere $\mbox{dist}(x,y)<R$, we would see
infinitely many disconnected pieces at $R \to \infty$. Following
the consideration we have used to prove the lemma, it is easy
to realise that the minimum number of bonds, which are necessary
to connect all these pieces together, exceeds infinitely many times
the number of interfacial sites inside such a sphere at $R \to \infty$. 
It implies that, on large enough scales, the interface is essentially
disconnected (further referred as frustrated) for any $d<2$
irrespective to that how small is $\epsilon = 2-d>0$.

Lattices with $1<d<2$ are, e.~g., Sierpi\'nski carpets
(see Fig.~2 in~\cite{Gefen}) which, however, do not
really have the interfacial properties of an analytic
continuation from $d$--dimensional hypercubes.
A crossection line in this case consists
of disconnected pieces, distributed in a fractal way,
as we have discussed already. However, if the interface would be
induced by appropriate boundary conditions,
then its fractal dimension would be $d_s<d-1$ rather than $d-1$,
since the minimum of free energy corresponds more or less to
the minimal crossection with $d_s<d-1$.
In this aspect, some random (statistical) lattices,
which cannot be split in a special way,
could be better candidates to mimic an analytic continuation
from integer $d$.

\section{Fluctuations of a frustrated interface at $d<2$}
\label{sec:fl}

Since the $(d-1)$--dimensional
interface becomes frustrated (disconnected) at $d<2$,
a formal analytic continuation from
$d$--dimensional hypercubes with $d \ge 2$ hardly can be
applied to describe it.
The disconnected pieces can relatively
freely move with respect to each other
within some range allowed by the lattice structure,
which is a qualitatively
new feature as compared to connected interfaces. One may expect
that it gives an extra contribution to the interfacial entropy.
The pieces of the frustrated interface typically has to be
located in such a way to make the narrowest connections between the
coexisting phases, i.~e., to minimize the free energy.
Therefore, on larger scales, the fluctuations in a random lattice
are expected to be jump--like, where the pieces of interface
are moved from one set of narrow places to another.
This interpretation becomes rather clear in a particular case
of randomized Sierpi\'nski carpets, obtained by cutting
out of the 2D lattice holes of different random shapes (starting
from larger holes, then, hierarchially, smaller and smaller holes).
The structure of such a lattice with suitable fractal
dimension $1<d<2$ consists of a set of holes with ``bridges'' in
between, when looking on any scale.
The places, where the pieces of the interface most probably
can be located, correspond to the narrowest crossections of
these ``bridges''.  We include a randomization,
since it eventually could be helpful to mimic essential properties
of $d$--dimensional hypercubes, as discussed at the end of
Sec.~\ref{sec:phin}.

We propose some phenomenological arguments to describe 
the above discussed fluctuations of a frustrated interface
in a random lattice.
In this consideration the fractal dimension of the interface
has to be $d_s<1$, but not necessarily $d-1$.
On a phenomenological level of description, one can introduce
a subset of lattice sites $\Omega$, where the interface most probably
can be located. It means that only the relevant spin configurations
are considered such that $\Lambda_S \subseteq \Omega$,
which correspond to local minima of free energy.
An essential quantity is the probability $Q(m)$ that a local
displacement of the interface by a distance $m$ is ``allowed''
by the lattice structure, i.~e., that it corresponds to
$\Lambda_S \to \Lambda'_S$ where $\Lambda_S, \Lambda'_S \subseteq \Omega$.
The displacement measured from $x \in \Lambda_S$ can be
defined as the minimal distance from $x$ to some $x' \in \Lambda'_S$,
i.~e., $m(x) = \inf\limits_{x'\in \Lambda'_S} \mbox{dist}(x,x')$.
The probability $Q(m)$ then is
$Q(m) = N_S^{-1} \sum\limits_{x \in \Lambda_S} I(m,x)$,
where $N_S$ is the number of elements in $\Lambda_S$ and
$I(m,x)$ is the indicator function. It has the value $I(m,x)=1$
if there exists $\Lambda'_S \subseteq \Omega$ such that
$\inf\limits_{x'\in \Lambda'_S} \mbox{dist}(x,x')=m$ holds,
and $I(m,x)=0$ otherwise.
Since the distribution of the interfacial sites around an
arbitrarily choosen $x \in \Lambda_S$
is characterised by certain fractal dimension $d_s$, the distribution
of the places $x' \in \Omega$, where the interface can be
eventually located, and hence the distribution of those values
of $m$ for which $I(m,x)=1$ holds also should be characterised by some
fractal dimension. Thus, the number of such values
of $m$ within $[0;M]$ has to increase like $M^{d'_s}$ at $M \to \infty$,
where $d'_s < 1$ is a fractal dimension.
Hence, the expected asymptotic behaviour of
$Q(m)$ for large $m$ is $Q(m) \sim m^{d'_s-1}$.
It is true for a frustrated interface at $d<2$.
To the contrary, $Q(m) \equiv 1$ corresponds to regular lattices
with $d =2,3,4, \ldots \,$,
where the displacements of the interface are quasy--continuous.

Let us assume that the interface is someway pinned at one its point
$x \in \Lambda_S$ (we may consider this as a constraint
for the spin configurations allowed) and consider the
probability distribution function $\mathcal{P}(m,L)$
over the displacements $m$ of the interface from its energetically most
preferable position, measured at some point $y \in \Lambda_S$
at a distance $L= \mbox{dist}(x,y)$ from $x$.
Considering the limit $L \to \infty$, it is suitable
to make an averaging over the set of lattice sites $y$ obeing the relation
$\mid \mbox{dist}(x,y) - L \mid < \varepsilon L$, where
$\varepsilon$ is small and positive. In this case the density
of the local minima of free energy is given by $Q(m)$.
A question arises whether or not the 
probability distribution over the ``allowed'' states,
which correspond to these local minima, can be characterised by
certain stiffness coefficient $\kappa_L$. If not, then it already means
that the fluctuations of the frustrated interface cannot be described
by a formula similar to~(\ref{eq:Zd}). If yes,
then an analogous formula reads
\begin{equation}
\bar \mathcal{P}(m,L) \sim
Q(m) \exp \left( - L^{d_s-2} \kappa_L m^2 \right) \;.
\label{eq:uh}
\end{equation}
It is supposed that first the normalized probability distribution
function $\mathcal{P}(m,L)$ is found for each individual $y$ and then
the distribution $\bar \mathcal{P}(m,L)$ is calculated
by an averaging over $y$. Besides, an $m$--independent prefactor
is omitted in~(\ref{eq:uh}). The validity of~(\ref{eq:uh})
is restricted to a region $L^{d_s-2} \kappa_L m^2 <C$,
where $C$ is some constant, to ensure that $\mathcal{P}(m,L)$
is not essentially influenced by relatively small variations
in the distance $\mbox{dist}(x,y)$. This equation agrees with~(\ref{eq:Zd})
at $Q(m) \equiv 1$. The latter relation can be valid for a
connected interface at $d \ge 2$.
However, as we have discussed already, $Q(m) \sim m^{d'_s-1}$ with
$d'_s<1$ is expected for large $m$ in our case of $d<2$ and $d_s<1$,
where the interface is necessarily disconnected or frustrated
(cf.~the \textit{Lemma}).
Hence, these arguments suggest that, in any case, the character
of the interfacial fluctuations changes qualitatively at $d<2$,
as compared to a formal analytic continuation from $d$--dimensional
hypercubes with $d \ge 2$. 
Thus, a formal extension of~(\ref{eq:cont})
to $d<2$, likely, has no physical meaning.

\section{Problem of an analytic continuation\\ of the critical exponents}

As shown in Secs.~\ref{sec:phin} and~\ref{sec:fl},
$d=2$ is a special lower marginal
value of $d$ as regards the behaviour of the interface between
the coexisting phases.
The bulk critical behaviour results from
a competition of large coexisting domains having opposite
sign of the mean magnetisation. Therefore the bulk critical behaviour
should be influenced by the interfacial structure.
Besides, the critical exponent of the surface tension
$\mu$ is related to the bulk critical exponents via
$\mu + \nu = 2 - \alpha = d \nu$~\cite{Baxter}.
Hence, one can expect that $2$ is a special marginal value
of $d$ also for these critical exponents. It would mean that an
analytic continuation of both interface and bulk critical exponents
from $d \ge 2$ (or $d=2,3$) to $d<2$
is only formal, like a continuation from $d<4$
to $d>4$ (above the upper critical dimension $d=4$).
In this case no appropriate family
of lattices could be found, providing the critical exponents
as (almost) continuous functions of $d$
in agreement with such an analytic continuation.

A family of fractal lattices, which allows to treat $d$
as a continuous parameter in exact recurrence relations,
has been considered in~\cite{ReMa}. However, these
lattices are not quite appropriate to mimic an analytic
continuation from $d$--dimensional hypercubes. A particular
problem is that the number of bonds connected to one lattice
site is not bounded in the thermodynamic limit (i.~e.,
the property~(i) is violated).

\section{Conclusions}

\begin{enumerate}
\item
The surface tension of rough interfaces in 2D and 3D Ising models has been
discussed. The known results have been completed by some original
calculations. In summary, a formula is given [Eq.~(\ref{eq:cont})],
which allows to interpolate the
corrections to finite--size scaling between two and three dimensions.
\item
It has been proven that $2$ is the marginal
value of $d$ below which the\linebreak $(d-1)$--dimensional
interface between the coexisting phases becomes
essentially disconnected or frustrated.
\item
Some phenomenological arguments have been proposed to describe
the fluctuations of such frustrated interfaces. They show
that $2$ is a special value of the dimension $d$ such that
the interfacial properties at $d<2$ disagree
with a formal analytic continuation from $d \ge 2$.
It, probably, is true also for the related interface and bulk
critical exponents.

\end{enumerate}

\end{document}